\documentclass[floatfix,twocolumn,showpacs,preprintnumbers,amsmath,amssymb,pra,superscriptaddress,longbibliography]{revtex4-1}
\usepackage{color}
\usepackage[usenames,dvipsnames,svgnames,table]{xcolor}
\usepackage[colorlinks=true,linkcolor=blue,urlcolor=blue,citecolor=blue]{hyperref}
\usepackage{mathtools}
\usepackage{graphicx}
\usepackage{dcolumn}
\usepackage{array}
\usepackage{lipsum}
\usepackage{bm}
\usepackage{subfigure}
\usepackage{amssymb}
\usepackage{multirow}
\usepackage{tabularx}
\usepackage{amsmath}
\usepackage{braket}
\graphicspath{{plots/}}
 \usepackage{lipsum}
\usepackage{mathrsfs}

\newcommand{\beq}{\begin{equation}}
\newcommand{\eeq}{\end{equation}}
\newcommand{\bea}{\begin{eqnarray}}
\newcommand{\eea}{\end{eqnarray}}

\begin{document}
\title{
Electronic pair alignment and roton feature in the warm dense electron gas
}

\author{Tobias Dornheim}
\email{t.dornheim@hzdr.de}

\affiliation{Center for Advanced Systems Understanding (CASUS), D-02826 G\"orlitz, Germany}
\affiliation{Helmholtz-Zentrum Dresden-Rossendorf (HZDR), D-01328 Dresden, Germany}

\author{Zhandos A.~Moldabekov}

\affiliation{Center for Advanced Systems Understanding (CASUS), D-02826 G\"orlitz, Germany}
\affiliation{Helmholtz-Zentrum Dresden-Rossendorf (HZDR), D-01328 Dresden, Germany}

\author{Jan Vorberger}
\affiliation{Helmholtz-Zentrum Dresden-Rossendorf (HZDR), D-01328 Dresden, Germany}

\author{Hanno K\"ahlert}

\affiliation{Institut f\"ur Theoretische Physik und Astrophysik, Christian-Albrechts-Universit\"at zu Kiel, D-24098 Kiel, Germany}

\author{Michael Bonitz}

\affiliation{Institut f\"ur Theoretische Physik und Astrophysik, Christian-Albrechts-Universit\"at zu Kiel, D-24098 Kiel, Germany}

\begin{abstract}
The study of matter under extreme densities and temperatures as they occur e.g. in astrophysical objects and nuclear fusion applications has emerged as one of the most active frontiers in physics, material science, and related disciplines. In this context, a key quantity is given by the dynamic structure factor $S(\mathbf{q},\omega)$, which is probed in scattering experiments---the most widely used method of diagnostics at these extreme conditions. In addition to its crucial importance for the study of warm dense matter, the modelling of such dynamic properties of correlated quantum many-body systems constitutes one of the most fundamental theoretical challenges of our time. Here we report a hitherto unexplained \emph{roton feature} in $S(\mathbf{q},\omega)$ of the warm dense electron gas, and introduce a microscopic explanation in terms of a new \emph{electronic pair alignment} model. This new paradigm will be highly important for the understanding of warm dense matter, and has a direct impact on the interpretation of scattering experiments. Moreover, we expect our results to give unprecedented insights into the dynamics of a number of correlated quantum many-body systems such as ultracold helium, dipolar supersolids, and bilayer heterostructures.
\end{abstract}


\maketitle

Matter at extreme densities and temperatures is ubiquitous throughout our universe~\cite{fortov_review} and naturally occurs in astrophysical objects such as giant planet interiors~\cite{Millot_Science_2015}, brown dwarfs~\cite{becker}, and  neutron stars~\cite{Haensel}. In addition, such \emph{warm dense matter} (WDM) conditions are highly relevant for cutting-edge technological applications such as the discovery of novel materials~\cite{Kraus2017,Lazicki2021}, hot-electron chemistry~\cite{Brongersma2015}, and inertial confinement fusion~\cite{hu_ICF,Betti2016}. Consequently, WDM is nowadays routinely realized in experiments in large research facilities around the globe such as the National Ignition Facility~\cite{Zylstra2022} in the USA, the European XFEL in Germany~\cite{Tschentscher_2017}, and SACLA~\cite{SACLA_2011} in Japan. 
Indeed, the advent of new experimental techniques for the study of WDM~\cite{falk_wdm} has facilitated a number of spectacular achievements~\cite{Fletcher2015,Kraus2016,Kraus2017,Knudson_Science_2015,Dias_Silvera_Science_2017} and has opened up new possibilities for the exciting field of \emph{laboratory astrophysics}.

\begin{figure*}\centering
\includegraphics[width=0.4275\textwidth]{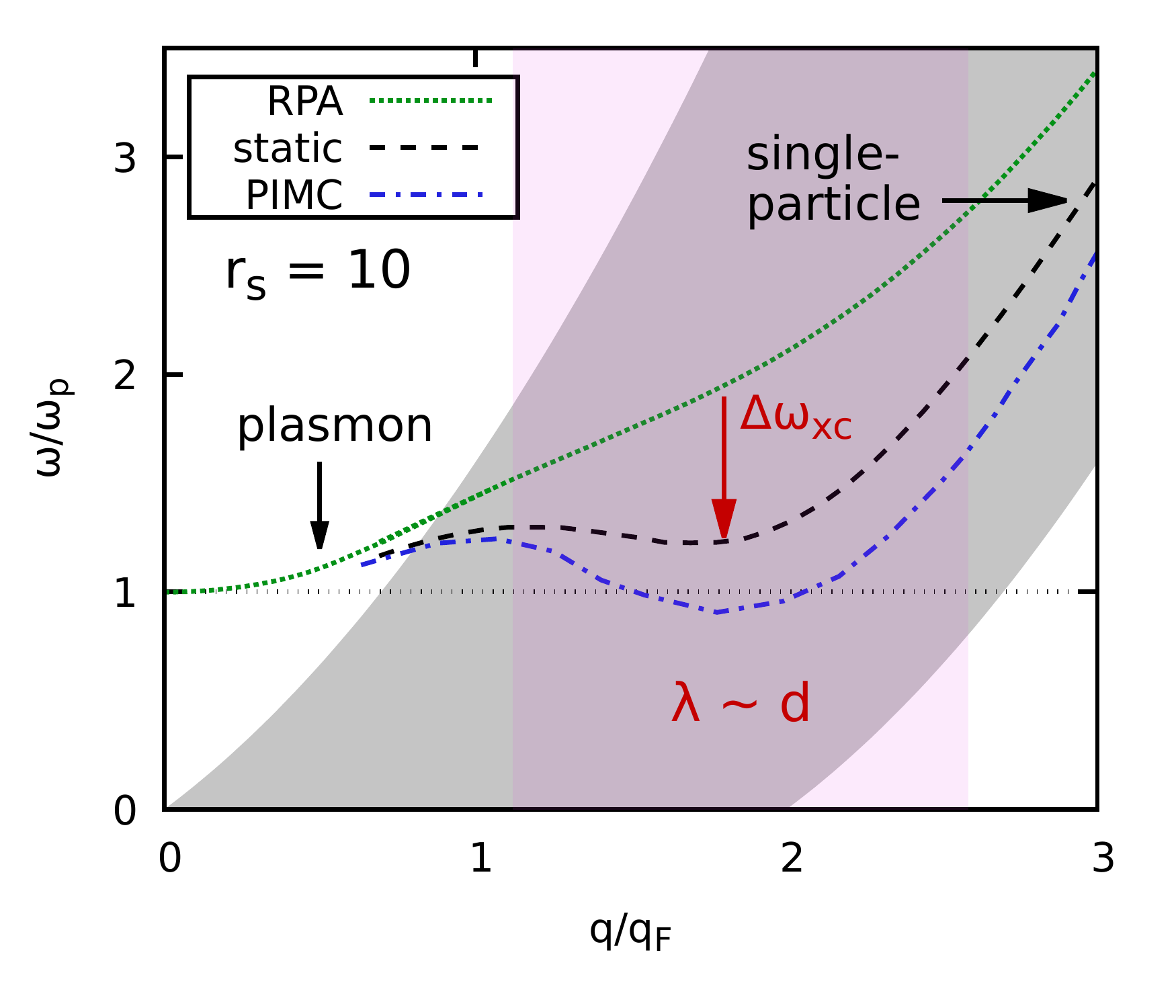}
\includegraphics[width=0.4275\textwidth]{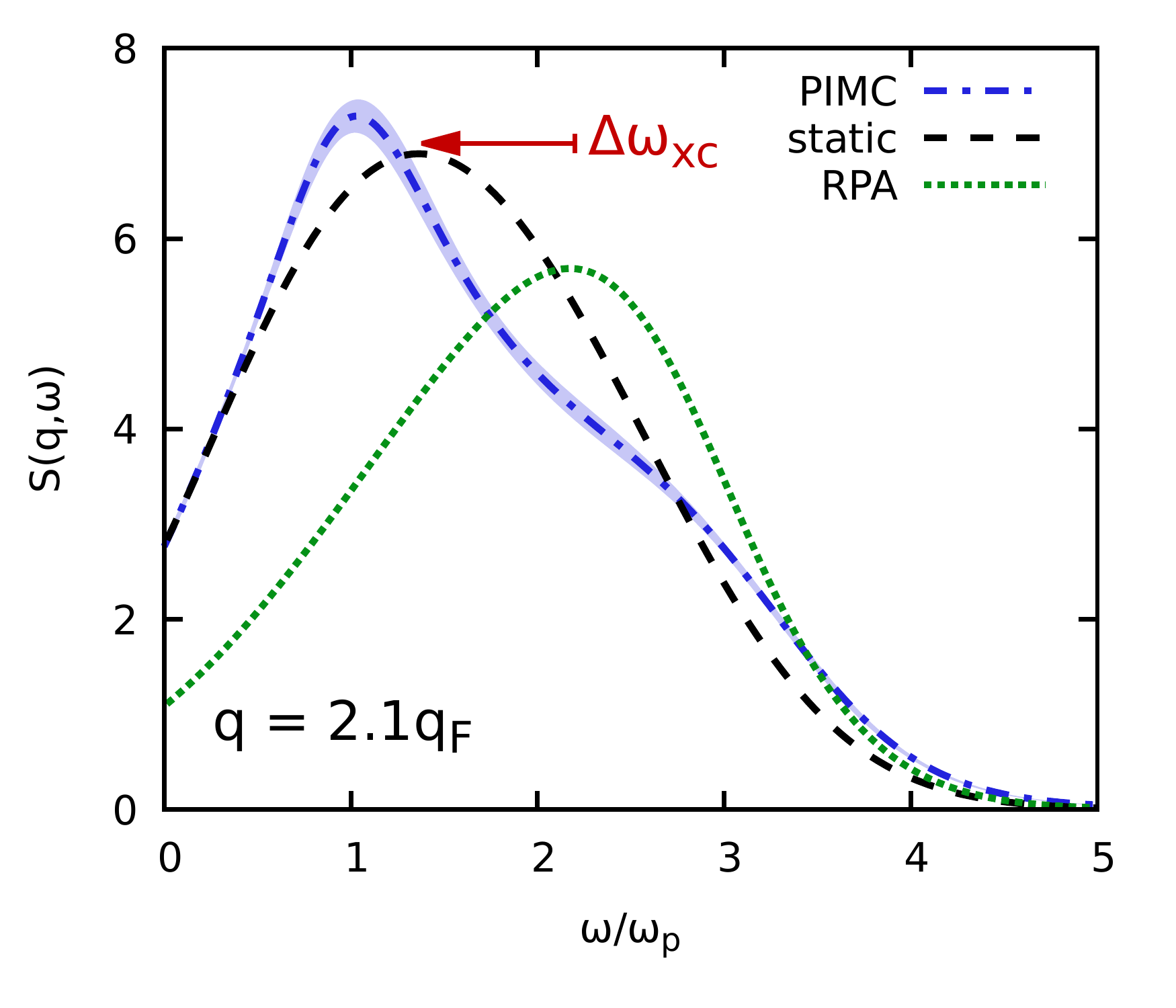}
\caption{\label{fig:pair}
Left: Spectrum of density fluctuations $\omega(q)$ in the uniform electron gas at the electronic Fermi temperature ($\theta=1$) at $r_s=10$. Dotted green: random phase approximation; dash-dotted blue: exact PIMC results~\cite{dornheim_dynamic}; dashed black: \emph{static approximation} $G(\mathbf{q},\omega)\approx G(\mathbf{q},0)$. For small wave numbers, the spectrum features a single sharp plasmon excitation. This collective regime, where the wave length $\lambda=2\pi/q$ is much larger than the average interparticle distance $d$, $\lambda \gg d$, is well described by the RPA. Upon entering the pair continuum (shaded grey), the DSF becomes substantially broadened. The regime with $\lambda\sim d$ (shaded red) features a hitherto unexplained pronounced red-shift $\Delta\omega_\textnormal{xc}$ compared to RPA, which eventually resembles the \emph{roton feature} known from ultracold helium~\cite{cep,Dornheim_SciRep_2022,Godfrin2012,cep,Kalman_2010}. Finally, the single-particle regime with $q\gg q_\textnormal{F}$ and $\lambda\ll d$ is dominated by a broad peak with $\omega(q)\sim q^2$. Right: $\omega$-dependence of the DSF $S(q,\omega)$ at $q\approx2.1q_\textnormal{F}$. The \emph{static approximation} entails an effective average over the less trivial structure of the full PIMC curve.
}
\end{figure*}

One of the central practical obstacles regarding the study of WDM is given by the lack of reliable diagnostics. The extreme conditions prevent the straightforward measurement even of basic system parameters like the electronic temperature, which have to be inferred indirectly from other observations. In this situation, the X-ray Thomson scattering (XRTS) technique~\cite{siegfried_review} has emerged as the de-facto standard method of diagnostics. In particular, an XRTS measurement gives one access to the dynamic structure factor (DSF) $S(\mathbf{q},\omega)$ describing the full spectrum of density fluctuations in the system. 
The task at hand is then to match the experimental observation with a suitable theoretical model, thereby inferring important system parameters like the electronic temperature $T$, density $n$, or charge state $Z$.
Yet, the rigorous theoretical modelling of WDM, in general, and of an XRTS signal, in particular, constitutes a most formidable challenge~\cite{wdm_book,review}. Indeed, the physical properties of WDM are characterized by the intriguingly intricate interplay of a number of effects such as the Coulomb interaction between charged electrons and ions, partial ionization and the formation of atoms and molecules, quantum effects like Pauli blocking and diffraction, and strong thermal excitations out of the ground state.

For this reason, the first rigorous results for the DSF of electrons in the WDM regime have been presented only recently~\cite{dornheim_dynamic} based on exact \emph{ab initio} path integral Monte Carlo (PIMC) simulations of the uniform electron gas (UEG)~\cite{review}. 
In particular, the UEG assumes a homogeneous neutralizing ionic background, and, therefore, allows us to exclusively focus on the rich effects inherent to the electrons. Due to its archetypical nature, the UEG constitutes one of the most fundamental model systems in physics, quantum chemistry, and related disciplines~\cite{quantum_theory}, and has been pivotal for a number of important developments, most notably the spectacular success of density functional theory in the description of real materials~\cite{Jones_RevModPhys_2015}. 
From a practical perspective, accurate results for the DSF of the UEG are indispensable for the interpretation of WDM experiments, and directly enter models such as the widely used Chihara decomposition~\cite{siegfried_review,Chihara_1987}.

While the availability of highly accurate results for $S(\mathbf{q},\omega)$ constitutes an important mile stone towards our understanding of the dynamics of correlated electronic matter, their theoretical interpretation has remained unclear. For example, the exact calculations by Dornheim \emph{et al.}~\cite{dornheim_dynamic} have uncovered a \emph{negative dispersion} in the UEG that closely resembles the \emph{roton feature} in quantum liquids such as $^4$He~\cite{cep} and $^3$He~\cite{Godfrin2012,Dornheim_SciRep_2022}. Despite speculations about a possible excitonic interpretation of this effect~\cite{Takada_PRL_2002,Takada_PRB_2016}, its precise nature is hitherto unknown. This reflects the notorious difficulty to describe the dynamics of correlated quantum many-body systems, which constitutes one of the most fundamental challenges in a number of research fields. In the present work, we introduce a new paradigm---the structural alignment of pairs of electrons---that allows us to understand, and to both qualitatively and quantitatively capture the \emph{roton feature} in the UEG. This breakthrough is of pivotal importance for the description of WDM and will have a direct and profound impact on a number of applications, such as the interpretation and design of nuclear fusion experiments~\cite{Moses_NIF}. Moreover, it will open up new avenues in a number of research fields and, in this way, will give novel insights into the dynamic behaviour of correlated quantum many-body systems such as ultracold helium~\cite{cep,Godfrin2012,Dornheim_SciRep_2022}, quantum dipole systems~\cite{Navon2021}, and strongly coupled bilayer heterostructures~\cite{Du2021}.

\textbf{Results.} In Fig.~\ref{fig:pair} (a), we show results for the corresponding spectrum of density fluctuations $\omega(q)$ that we estimate from the maximum in the DSF at the electronic Fermi temperature $\theta=k_\textnormal{B}T/E_\textnormal{F}=1$ (with $E_\textnormal{F}$ being the Fermi energy) and the density parameter $r_s=\overline{a}/a_\textnormal{B}=10$ (with $\overline{a}$ being the average distance to the nearest neighbour and $a_\textnormal{B}$ the first Bohr radius). The dotted green curve shows the ubiquitous random phase approximation (RPA), which entails a mean-field description of the electronic density response to an external perturbation; see the Supplemental Material for details. The dash-dotted blue curve shows exact PIMC results that have been obtained on the basis of the full frequency-dependent local field correction $G(\mathbf{q},\omega)$, which contains the complete wave-vector and frequency resolved information about electronic exchange--correlation effects. Finally, the dashed black curve corresponds to the \emph{static approximation}, i.e., by setting $G(\mathbf{q},\omega)=G(\mathbf{q},0)$; see Ref.~\cite{dornheim_dynamic} for a detailed explanation of the PIMC calculations.

Let us next discuss the different physical regimes shown in Fig.~\ref{fig:pair}.
For small $q$, i.e., in the collective regime where the wavelength is much larger than the average interparticle distance ($\lambda\gg d\sim2r_s$) the spectrum consists of a single, sharp plasmon peak that is exactly captured by all three theories. Upon increasing $q$, we enter the pair continuum, where the plasmon decays into a multitude of other excitations and ceases to be a sharp feature~\cite{quantum_theory}.  From a comparison to the simulations, it is evident that the RPA breaks down in this regime and does not capture the intriguing non-monotonous behaviour of the exact PIMC data. Indeed, the latter exhibit a pronounced minimum in $\omega(q)$ around $q=2q_\textnormal{F}$, which closely resembles the well-known \emph{roton feature} in both $^4$He~\cite{cep,griffin1993excitations} and $^3$He~\cite{Godfrin2012,Dornheim_SciRep_2022}.  We stress that this is a real physical trend, which has been observed experimentally for electrons in alkali metals~\cite{vomFelde_PRB_1989,Takada_PRL_2002}. In the present work, we demonstrate that this red-shift $\omega_\textnormal{xc}$ compared to RPA is a direct consequence of the \emph{alignment of pairs of electrons} where $\lambda\sim d$, and show that it can be understood and accurately quantified in terms of the microscopic spatial structure of the system. 
Finally, a further increase of $q$ eventually brings us into the single-particle regime ($\lambda\ll d$), where $\omega(q)$ is known to increase quadratically with $q$.

Evidently, the \emph{static approximation} leads to a substantial improvement over the RPA, and qualitatively reproduces both the pronounced red-shift and even exhibits a shallow minimum in $\omega(q)$ at the correct position. A more detailed insight is given in Fig.~\ref{fig:pair} (b), where we show the full $\omega$-dependence of $S(\mathbf{q},\omega)$ in the vicinity of the roton feature. The exact PIMC curve shows a nontrivial shape consisting of a pronounced peak at $\omega(q)$ and an additional shoulder around $\omega_\textnormal{RPA}(q)$. In contrast, the dashed black curve features a single broad peak that is located between both aforementioned features. In fact, the \emph{static approximation} can be understood as a kind of frequency-averaged description of the actual spectrum of density fluctuations, and, therefore, reproduces frequency-averaged properties like the static structure factor $S(q)$~\cite{Dornheim_PRL_2020_ESA} with remarkable accuracy. 
Moreover, it does capture the correlation-induced shift in $S(\mathbf{q},\omega)$ towards lower frequencies, which is the root cause of the \emph{roton feature} in the UEG that is studied in the present work.

\begin{figure*}\centering
\includegraphics[width=0.475\textwidth]{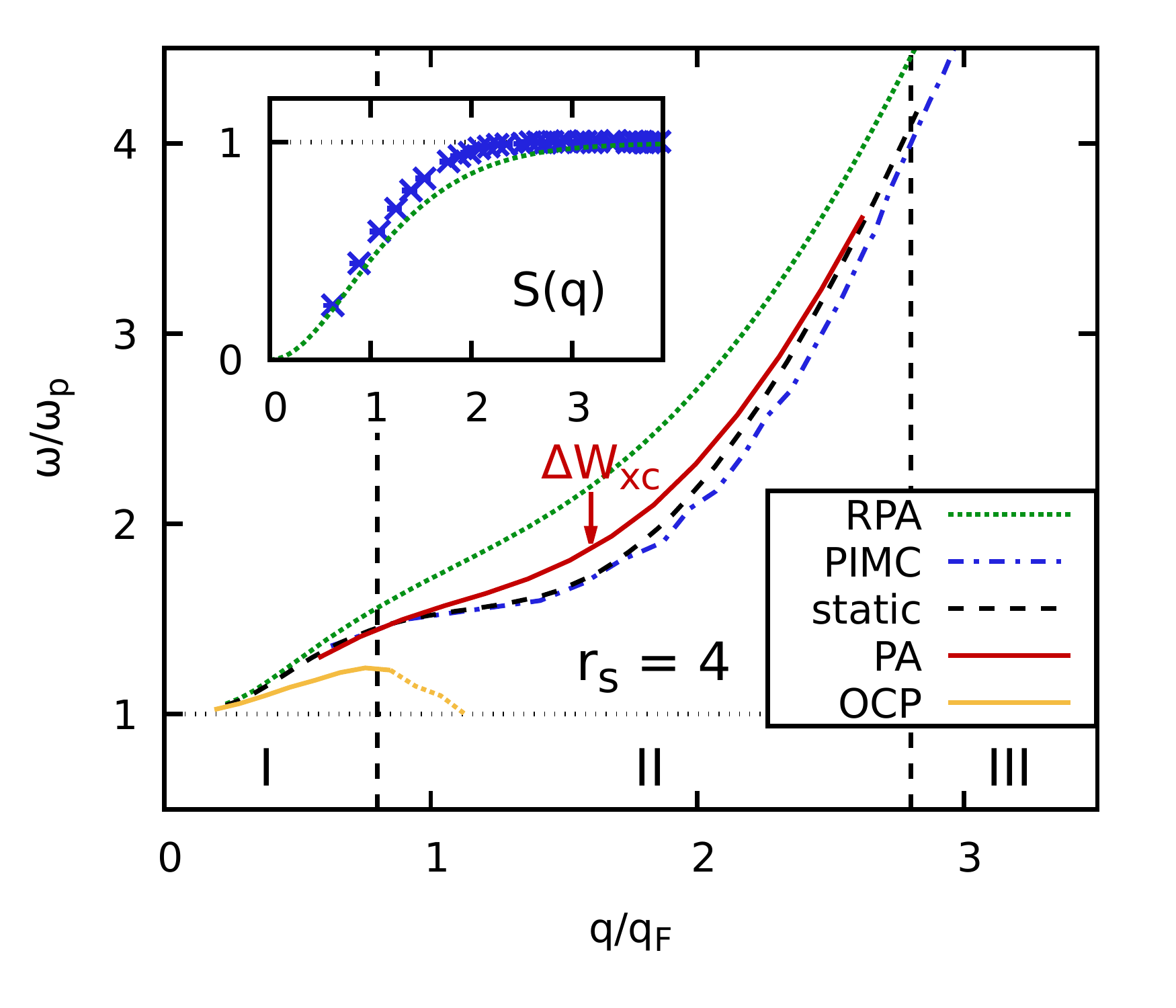}\includegraphics[width=0.475\textwidth]{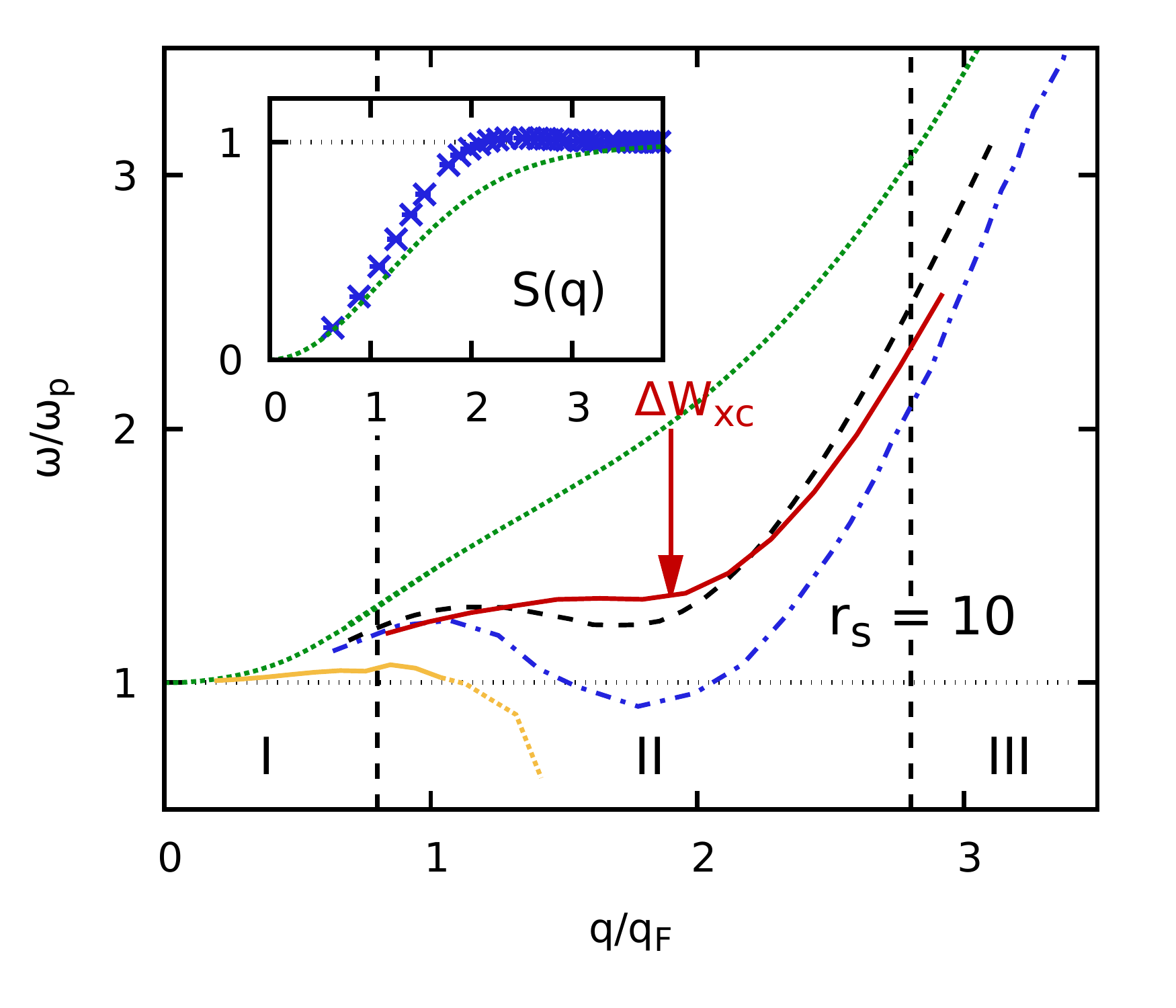}
\caption{\label{fig:dispersion}
Spectrum of density fluctuations [peak of $S(\mathbf{q},\omega)$] in the warm dense electron gas at the electronic Fermi temperature for $r_s=4$ (left) and $r_s=10$ (right). Dotted green: RPA; dash-dotted blue: exact PIMC results~\cite{dornheim_dynamic} using full dynamic LFC $G(\mathbf{q},\omega)$; dashed black: \emph{static approximation}, i.e., setting $G^\textnormal{static}(\mathbf{q},\omega)=G(\mathbf{q},0)$; solid yellow: classical OCP at $\Gamma=1.92$ ($r_s=4$) and at $\Gamma=4.8$ ($r_s=10$); solid red: new \emph{electronic pair alignment} model [Eq.~(\ref{eq:dispersion})] introduced in this work;
the red arrows indicate the corresponding average red-shift compared to RPA.
We can distinguish 3 distinct physical regimes: I) $\lambda \gg d$ [collective], II) $\lambda\sim d$ [electronic pair alignment, two-particle excitations], and III) $\lambda \ll d$ [single-particle]. The insets show PIMC and RPA results for the static structure factor $S(q)$ and illustrate the absence of spatial structure at these conditions. 
}
\end{figure*}

To understand the physical origin of the latter, 
it is well-worth to explore possible analogies to other systems, most notably $^4$He~\cite{cep,griffin1993excitations} and $^3$He~\cite{Godfrin2012,Dornheim_SciRep_2022}. In addition, we mention the extensively explored negative dispersion in the classical one-component plasma~(OCP)~\cite{Mithen_OCP_2012}. Remarkably, both cases have been explained by the onset of spatial localization of the particles~\cite{Kalman_2010}, and the roton feature can then be quantified in terms of $S(\mathbf{q})$, e.g.~via the Feynman ansatz~\cite{Feynman_ansatz} for He. 
In stark contrast, such structural arguments do not apply to the present case of the warm dense UEG. 
Indeed, the maximum in $S(q)$ does not exceed $1.02$ even at $r_s=10$ [see the inset of Fig.~\ref{fig:dispersion} (b)] and the system is largely disordered.

\begin{figure}\centering
\includegraphics[width=0.475\textwidth]{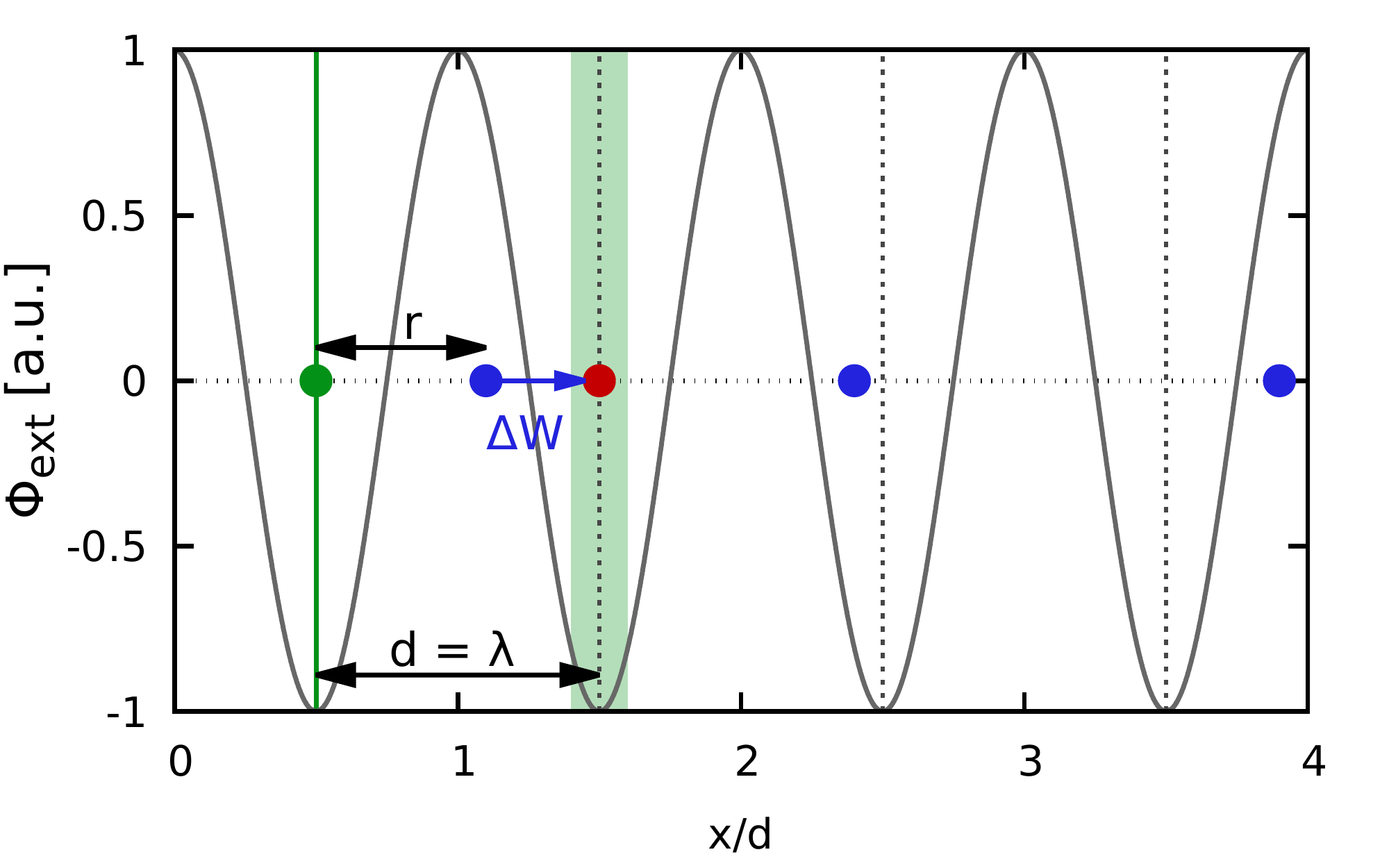}
\caption{\label{fig:actual}
Illustration of the \emph{alignment of electron pairs}: Let the green bead be a fixed reference particle. Without an external perturbation [$\phi_\textnormal{ext}$], the system is disordered on average in the WDM regime, see the blue beads. In order to follow $\phi_\textnormal{ext}$, the particles have to re-align themselves to the minima of the latter, see blue arrow + red bead. 
In the process, they change the potential energy of the green reference particle by an amount of $\Delta W$, see Eq.~(\ref{eq:shift}). In the regime of electronic pair alignment, $\lambda\sim d$, the energy shift $\Delta W$ is substantially negative. In other words, a density fluctuation with a corresponding $q=2\pi/\lambda$ contains comparably less energy due to its alignment to the potential energy landscape of the system (shaded green area).
RPA substantially underestimates this effect and, therefore, does not capture this correlation-induced \emph{red-shift}. 
}
\end{figure} 

\begin{figure*}[t!]\centering
\includegraphics[width=0.475\textwidth]{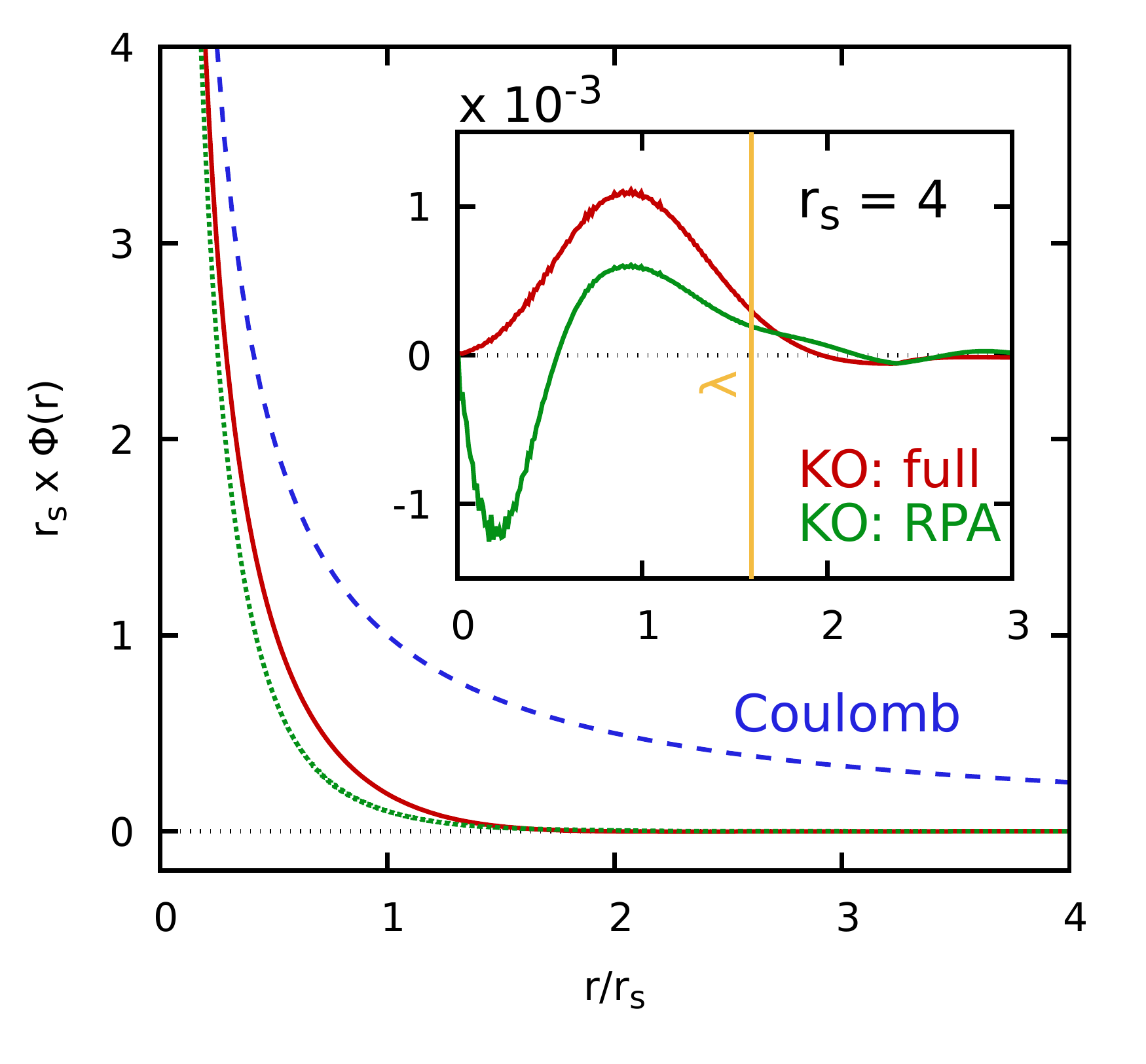}\includegraphics[width=0.475\textwidth]{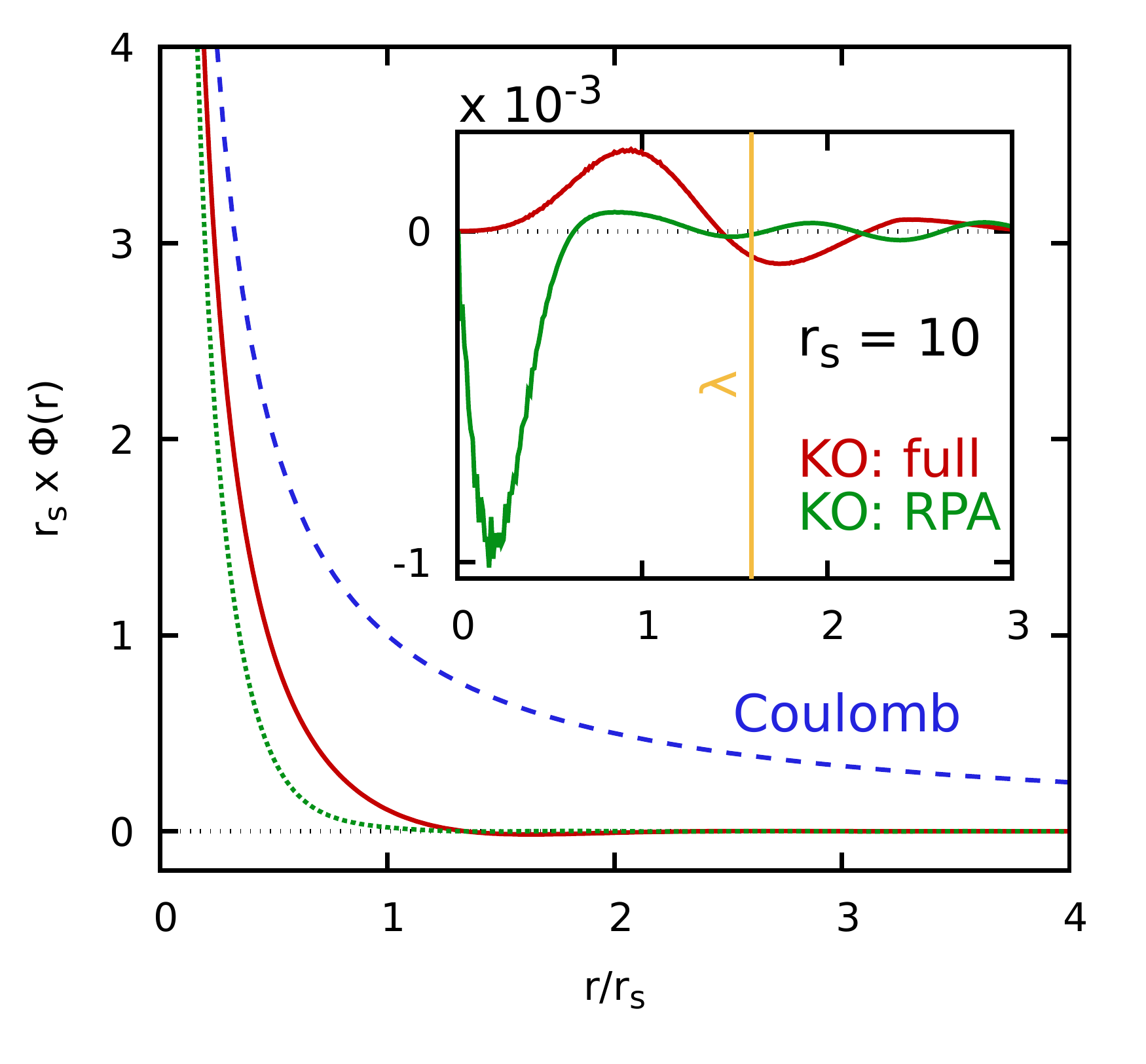}
\caption{\label{fig:KO}
Effective Kukkonen Overhauser (KO) potential between a pair of electrons surrounded by the electronic medium. Dotted green: RPA (mean field); solid red: full KO potential using exact PIMC data for the LFC $G(\mathbf{q},0)$~\cite{dornheim_ML}; dashed blue: bare Coulomb. 
The inset shows the contribution to the full shift $\Delta W(q)$ for $q\approx 2q_\textnormal{F}$ within LFC and RPA as a function of the distance between the particles $r$. For $r_s=4$, positive and negative contributions to $\Delta W(q)$ approximately cancel within the RPA, and the red-shift in $\omega(q)$ is mainly due to the reduction of the interaction energy described by the exact PIMC results. For $r_s=10$, the RPA even predicts an unphysical increase of $W$, whereas our PIMC results again correctly describe the minimization of the interaction energy for $\lambda\sim d$.
}
\end{figure*}

To explain the physical mechanism behind the red-shift and eventual roton feature in the warm dense UEG, we explore the nature of the excitations of density fluctuations in this regime in Fig.~\ref{fig:actual}. More specifically, the green bead depicts an arbitrary reference particle, and the blue beads other electrons which are, on average, disordered; this can be seen by the absence of pronounced features in $S(q)$. From a mathematical perspective, the dynamic structure factor entails the same information as the density response function that describes the response to an external harmonic perturbation~\cite{quantum_theory}. The latter is depicted by the black sinusoidal line and induces the leftmost blue particle to follow the perturbation, i.e., the blue arrow. Naturally, this reaction of the system is associated with a change in the potential energy by an amount $\Delta W$. In the case of $\lambda\sim d$, the particles will actually align themselves to the minima of the effective potential energy (shaded green area), which leads to a \emph{lowering} of the interaction energy compared to the unperturbed case. Equivalently, we can say that a density fluctuation contains comparably less energy when $\lambda\sim d$ as it coincides with a spatial pattern that minimizes the potential energy landscape.
This \emph{electronic pair alignment} is highly sensitive to an accurate treatment of electronic XC-effects and becomes more pronounced with increasing $r_s$.
We can thus express the exact spectrum of density fluctuations as
\begin{eqnarray}\label{eq:dispersion}
\omega(q) &=& \omega_\textnormal{RPA}(q) -  \Delta \omega_\textnormal{xc}(q)\ , \\
&=& \omega_\textnormal{RPA}(q) -  \alpha(q) \Delta W_\textnormal{xc}(q)\ , \nonumber
\end{eqnarray}
where we have assumed in the second line that the kinetic contribution to $\omega(q)$ is accurately treated within RPA. The corresponding absence of XC-effects onto the momentum distribution $n(q)$ is demonstrated in the Supplemental Material. 
The screening coefficient~\cite{kugler1} $\alpha(q) = \chi(q)/\chi_0(q)$ is given by the ratio of the full and noninteracting static density response functions and takes into account the fact that, on the static level, the UEG does not react to an external perturbation in the limit of $\lambda \gg d$. A more detailed discussion of the role of $\alpha(q)$ in our model is given below.
The exchange--correlation correction to the potential part of the excitation energy of a density fluctuation of wavenumber $q$ is given by
\begin{eqnarray}
\Delta W_\textnormal{xc}(q) &=& \Delta W(q) - \Delta W_\textnormal{RPA}(q)\ ,
\end{eqnarray}
where $\Delta W(q)$ denotes the actual change in the interaction energy. 
Eq.~(\ref{eq:dispersion}) implies that the observed red-shift in $\omega(q)$ is a direct consequence of the insufficient description of the \emph{electronic pair alignment} within RPA.  
To quantitatively evaluate this effect, we express the energy shift $\Delta W$ as
\begin{eqnarray}\label{eq:shift}
\Delta W(q) =  n  \int \textnormal{d}\mathbf{r}\ g(r) \left[  \phi(r) - \phi(r_{q}) \right] \ ,
\end{eqnarray} where $\phi(r)$ denotes the effective potential between two electrons in the presence of the electron gas.
From a physical perspective, Eq.~(\ref{eq:shift}) can be interpreted as follows: a reference particle at $\mathbf{r}=\mathbf{0}$ is located in the minimum of an external sinusoidal perturbation, cf.~Fig.~\ref{fig:actual}. On average, it will encounter a second particle at $\mathbf{r}$ with the probability $P(\mathbf{r})=n g(r)$, where $g(r)$ is the usual radial distribution function. Finally, we have to evaluate the difference in the effective potential between $\mathbf{r}$, and the position of the nearest minimum in the external potential, which we denote $r_{q}$.  Without loss of generality, we assume a perturbation along the $\mathbf{x}$-direction. 
For $\lambda\gg d$, this physical picture breaks down as the translation of the second particle to $r_q$ will be increasingly blocked by other electrons in the system. This is a direct manifestation of screening in our model, and is taken into account by the coefficient $\alpha(q)$ in Eq.~(\ref{eq:dispersion}).
A relation of the energy shift Eq.~(\ref{eq:shift}) to the XC-contribution to the self energy known from Green functions theory is given in the Supplemental Material.

The appropriate effective potential between two electrons has to 1) take into account all effects of the surrounding medium and 2) not include any XC-effects between the electrons themselves. This is a crucial difference to the effective interaction of two test charges in a medium, where one can simply use dielectric theories~\cite{quantum_theory}. In the present context, the appropriate potential has been derived by Kukkonen and Overhauser~\cite{Kukkonen_Overhauser_PRB_1979} (KO), and is given by a functional of the LFC, $\phi(r)=\phi[G(\mathbf{q},0)](r)$. Here we use exact PIMC results~\cite{dornheim_ML} for $G(\mathbf{q},0)$ and perform a Fourier transform to obtain the corresponding KO potential $\phi(r)$. The results are shown in Fig.~\ref{fig:KO} for the metallic density of $r_s=4$ (left) and the more strongly coupled case of $r_s=10$ (right), with the solid red, dotted green, and dashed blue lines depicting the KO potential with the LFC, the KO potential in RPA, and the bare Coulomb potential, respectively. Evidently, the impact of the medium vanishes for $r\to0$, and all curves converge. In addition, both KO potentials quickly decay for $r\gtrsim 2 r_s$ and do not exhibit the long Coulombic tail.

The insets in Fig.~\ref{fig:KO} show the respective contributions to $\Delta W(2q_\textnormal{F})$ [Eq.~(\ref{eq:shift})], and the vertical yellow line depicts the corresponding wave length $\lambda$. For $r_s=4$, the positive and negative contributions to $\Delta W_\textnormal{RPA}$ nearly cancel. Consequently, the observed XC-induced red-shift in Fig.~\ref{fig:dispersion} (a) is predominantly due to the lowering of the interaction energy, i.e., $\Delta W(q)$, in the regime of \emph{electronic pair alignment}. For $r_s=10$, the situation is more subtle, and the RPA predicts a substantial increase in $\omega(q)$ for $q\sim2q_\textnormal{F}$. This is a direct consequence of the pair correlation function $g_\textnormal{RPA}(r)$, which is known to be strongly negative for small $r$ at these conditions. The exact PIMC results indicate a similar trend as for the metallic density and again indicate a lowering of $W$ due to the \emph{electronic pair alignment}. Therefore, it is the combination of 1) removing the spurious RPA prediction for $W$ and 2) further adding the correct decrease in $W$ quantified by our PIMC simulations that leads to the large down-shift of the actual $\omega(q)$ compared to RPA.

Let us now apply these new insights to the spectrum of density fluctuations depicted in Fig.~\ref{fig:dispersion}. Specifically, the solid red lines show our new model Eq.~(\ref{eq:dispersion}), which reproduces the correct behaviour of $\omega(q)$ at both densities. We note that it follows the \emph{static approximation} rather than the full dynamic PIMC data at $r_s=10$. This is expected, as Eq.~(\ref{eq:shift}) constitutes an average over changes in the effective potential $\phi$ for different initial positions $r$. Therefore, it gives us the \emph{average change} in $\omega(q)$, i.e., the location of the peak of the broad dashed black curve in Fig.~\ref{fig:pair}, and not the actual position of the sharper roton peak. The correctness of our model is demonstrated over a broad range of densities and temperatures in the Supplemental Material.

In combination, our new results provide a complete description of $\omega(q)$ over the full $q$-range, and allow us to give a simple and physically intuitive explanation of both the XC-induced red-shift at metallic density and the \emph{roton feature} at stronger coupling. For small $q$, the main feature of $\omega(q)$ is given by the sharp plasmon peak. In particular, the plasmon is a \emph{collective excitation} and involves all particles in the system. Upon entering the pair continuum, the DSF broadens, and we find an initial increase of $\omega(q)$ with $q$; this is a well-known kinetic effect due to quantum delocalization and is qualitatively reproduced by the RPA. Correspondingly, it is not present in $\omega(q)$ of the classical OCP (yellow curves in Fig.~\ref{fig:dispersion}) at the same conditions. 
In the vicinity of $q_\textnormal{F}$, the potential contribution to $\omega(q)$ starts to be shaped by the \emph{electronic pair alignment}, and the corresponding lowering of the interaction energy $W$ leads to the observed red-shift. 
In other words, the non monotonic \emph{roton feature} at $r_s=10$ is a consequence of two competing trends: 1) the delocalization-induced quadratic increase in $\omega(q)$ and 2) the decrease of the interaction contribution due to \emph{electronic pair alignment}.

An additional insight into the physical origin of the excitations in this regime comes from the effective potential shown in Fig.~\ref{fig:KO}. Specifically, $\phi(r)$ vanishes for $r\gtrsim2r_s$, which means that the change in the interaction energy $W$ is of a remarkably local nature. This, in turn, strongly implies that the \emph{roton feature} is due to two-particle excitations---the \emph{alignment of electron pairs}. In this light, we can even explain the nature of the nontrivial structure of the full $S(q,\omega)$ shown in Fig.~\ref{fig:pair} (b): the large $\Delta W$ leading to the actual roton peak in the blue curve must be a result of configurations where two particles are initially separated by a short distance $r < r_s$. Only in this case will the corresponding change in $\phi(r)$ be sufficient for the observed lowering in $\omega(q)$. The additional shoulder in $S(q,\omega)$ is then due to transitions where the particles have even in their initial configuration been effectively separated, so that the change in $\phi(r)$ and, hence, the resulting $\Delta W(q)$ are comparably small.

Going back to Fig.~\ref{fig:dispersion}, we note that a further increase in $q$ again changes the nature of the excitations. In particular, $\omega(q)$ is exclusively shaped by single-particle effects when $\lambda\ll r_s$.

\textbf{Discussion.} We are convinced that our new findings for the spectrum of density fluctuations of the UEG---one of the most fundamental quantum systems in the literature---will open up many new avenues for future research in a diverse array of fields. First and foremost, the archetypical nature of the UEG has allowed us to clearly isolate the rich interplay of the correlated electrons with each other, which constitutes an indispensable basis for the study of realistic systems. The \emph{roton feature} has already been experimentally observed for electrons in metals at ambient conditions~\cite{vomFelde_PRB_1989,Takada_PRL_2002}, and understanding the interplay of the \emph{electronic pair alignment} with the presence of the ions will be an important next step. We expect that the observation of a non-monotonous $\omega(q)$ will also be possible in the WDM regime for real materials such as hydrogen, since the presence of bound states leads to an effectively reduced electronic density~\cite{kremp_book} and, therefore, an increased effective Wigner-Seitz radius $r_s^*$.

Our new microscopic theory for the DSF in the regime of $\lambda\sim d$ will be particularly relevant for the interpretation of XRTS measurements~\cite{siegfried_review}, which constitute the key diagnostics of state-of-the-art WDM experiments.

Going beyond the study of WDM, we stress that the proposed concept of \emph{electronic pair alignment} is very general, and will likely help to shed light on the mechanism behind the spectrum of density fluctuations and elementary excitations in a number of correlated quantum systems. This includes the improved understanding of the "original" roton mode in liquid $^4$He~\cite{cep,Nozieres2004} and $^3$He~\cite{Godfrin2012,Dornheim_SciRep_2022}, the impact of supersolidity~\cite{Norcia2021,Saccani_Supersolid_PRL_2012} onto the DSF of strongly coupled dipole systems~\cite{Navon2021}, and the transition from the liquid regime to a highly ordered Wigner crystal~\cite{Zhou2021} in strongly coupled bilayer heterostructures~\cite{Du2021}.

\appendix
\section*{Supplemental Material}

\subsection*{Dynamic structure factor and density response}\label{sec:supplement} 

 The dynamic structure factor $S(\mathbf{q},\omega)$ is directly connected to the linear density response function by the well-known fluctuation--dissipation theorem~\cite{quantum_theory},
\begin{eqnarray}\label{eq:FDT}
S(\mathbf{q},\omega) = - \frac{\textnormal{Im}\chi(\mathbf{q},\omega)}{\pi n (1-e^{-\beta\omega})}\ .
\end{eqnarray}
It is very convenient to express the latter as
\begin{eqnarray}\label{eq:chi}
\chi(\mathbf{q},\omega) = \frac{\chi_0(\mathbf{q},\omega)}{1-\frac{4\pi}{q^2}\left[1-G(\mathbf{q},\omega)\right]\chi_0(\mathbf{q},\omega)}\ ,
\end{eqnarray}
where $\chi_0(\mathbf{q},\omega)$ describes the density response of a noninteracting system at the same conditions and can be readily computed.
As we have mentioned in the main text, the dynamic LFC~\cite{kugler1} $G(\mathbf{q},\omega)$ contains all electronic XC-effects; setting $G(\mathbf{q},\omega)\equiv0$ thus corresponds to the RPA. The DSF within RPA, the \emph{static approximation} $G^\textnormal{static}(\mathbf{q},\omega)\equiv G(\mathbf{q},0)$, or using the full PIMC results for $G(\mathbf{q},\omega)$~\cite{dornheim_dynamic} are then obtained by inserting the corresponding $\chi(\mathbf{q},\omega)$ into Eq.~(\ref{eq:FDT}). The DSF of the classical OCP is obtained from molecular dynamics simulations using the LAMMPS code~\cite{plimpton1995jcp}.

\subsection*{Definition of the effective potential}\label{sec:supplement2} 

The effective potential described in the main text has been first derived by Kukkonen and Overhauser~\cite{Kukkonen_Overhauser_PRB_1979}, and is given by
\begin{eqnarray}\label{eq:KO}
\Phi^\textnormal{KO}(\mathbf{q}) = \frac{4\pi}{q^2} + \left[\frac{4\pi}{q^2}\left(1-G(\mathbf{q},0)\right)\right]^2 \chi(\mathbf{q},0) \ ;
\end{eqnarray}
see also the excellent discussion by Giuliani and Vignale~\cite{quantum_theory}. Evidently, it is exclusively a functional of the LFC, and, therefore, highly sensitive to electronic XC-effects. In the present work, we use the highly accurate neural net representation of $G(q,\omega=0;r_s,\theta)$ by Dornheim \emph{et al.}~\cite{dornheim_ML} that is based on exact PIMC simulation data. The results for $\phi(r)$ in coordinate space are then obtained via a simple one-dimensional Fourier transform, which we solve numerically.

\begin{figure}\centering
\includegraphics[width=0.475\textwidth]{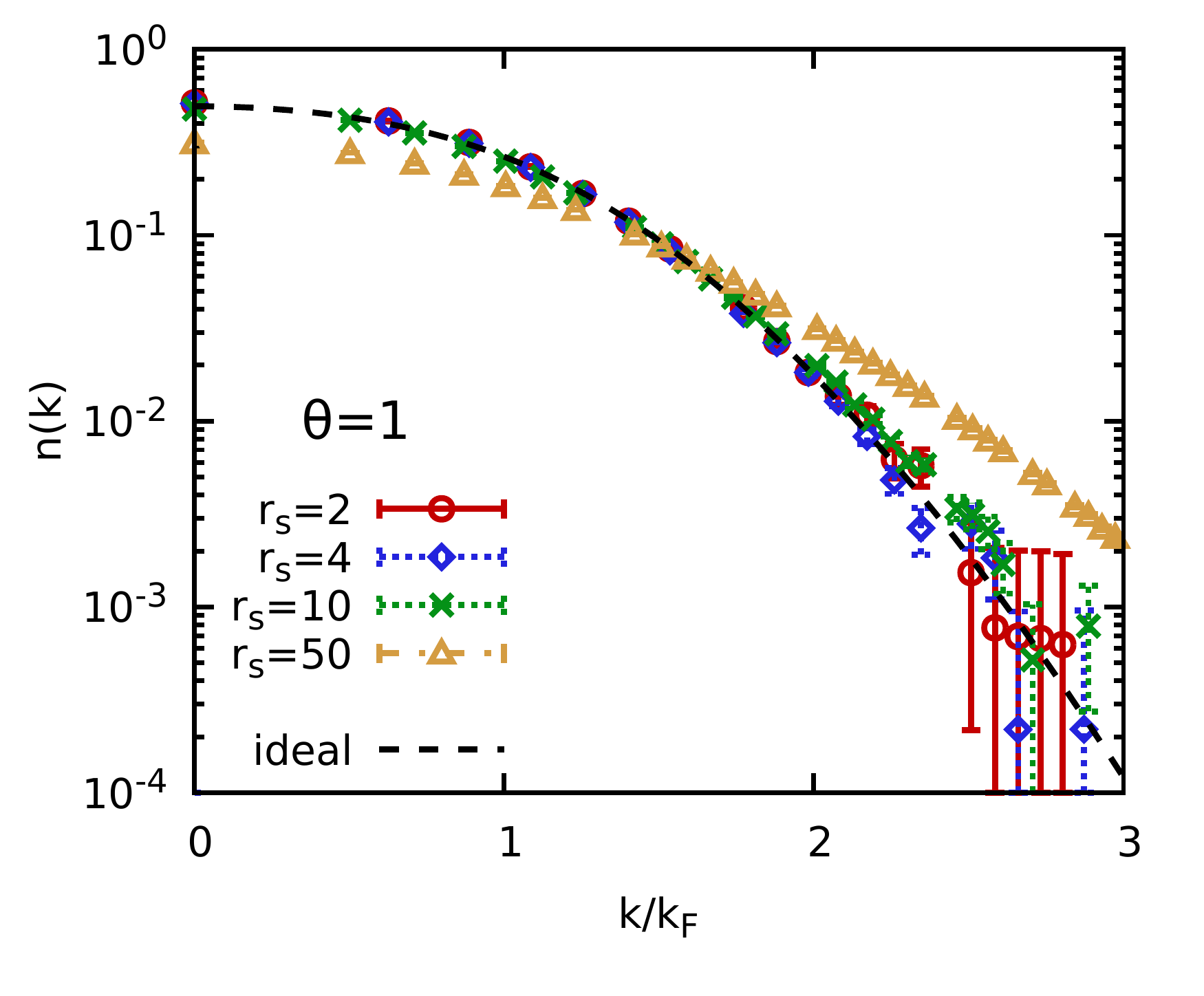}
\caption{\label{fig:single-particle}
Momentum distribution $n(k)$ of the uniform electron gas at the electronic Fermi temperature $\theta=1$. The symbols show exact PIMC results for different values of the Wigner-Seitz radius $r_s$, and the dashed black line the Fermi distribution function describing a noninteracting Fermi gas. Taken from Ref.~\cite{Dornheim_PRB_nk_2021}.
}
\end{figure}

\begin{figure*}\centering
\includegraphics[width=0.475\textwidth]{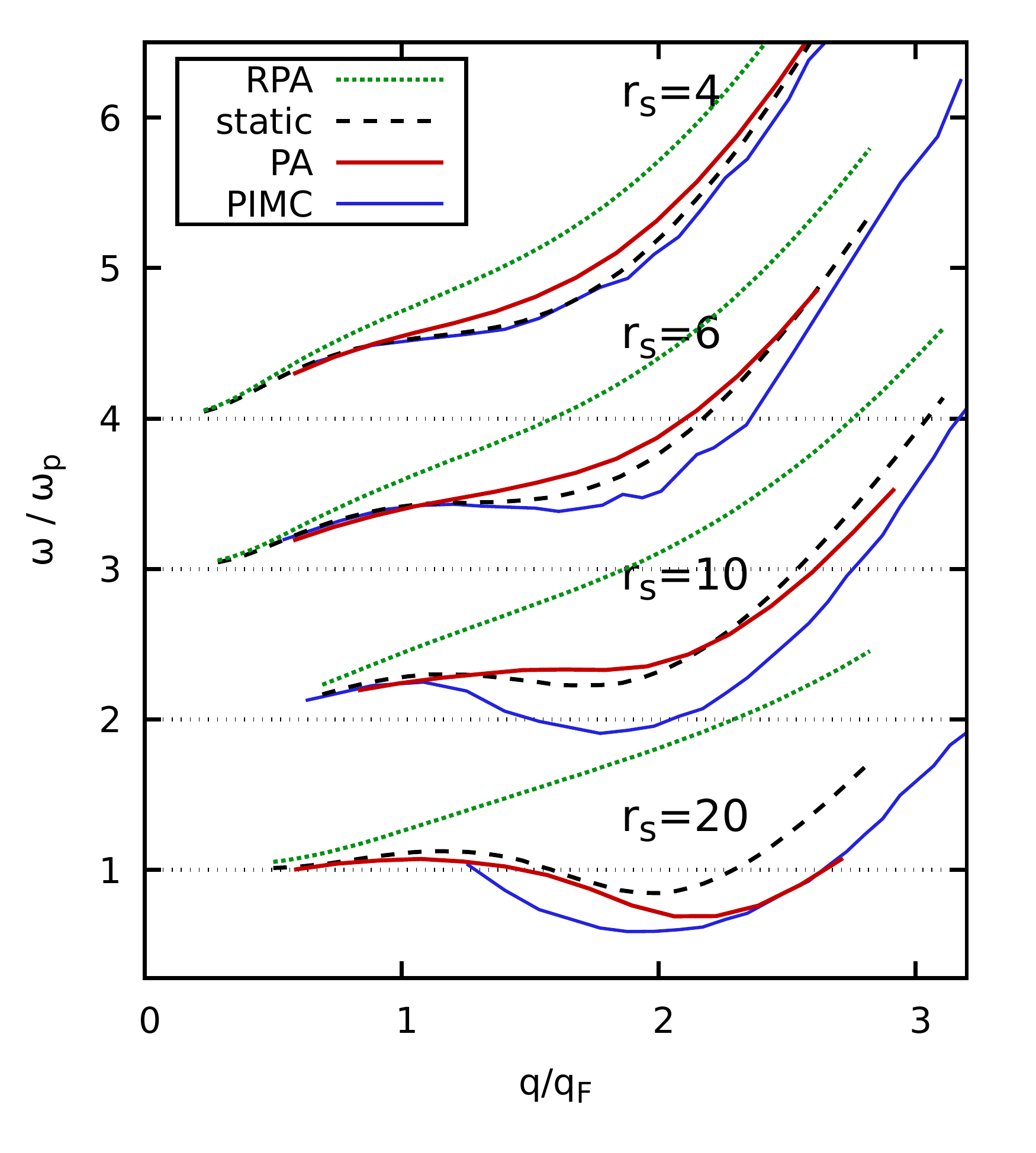}\includegraphics[width=0.475\textwidth]{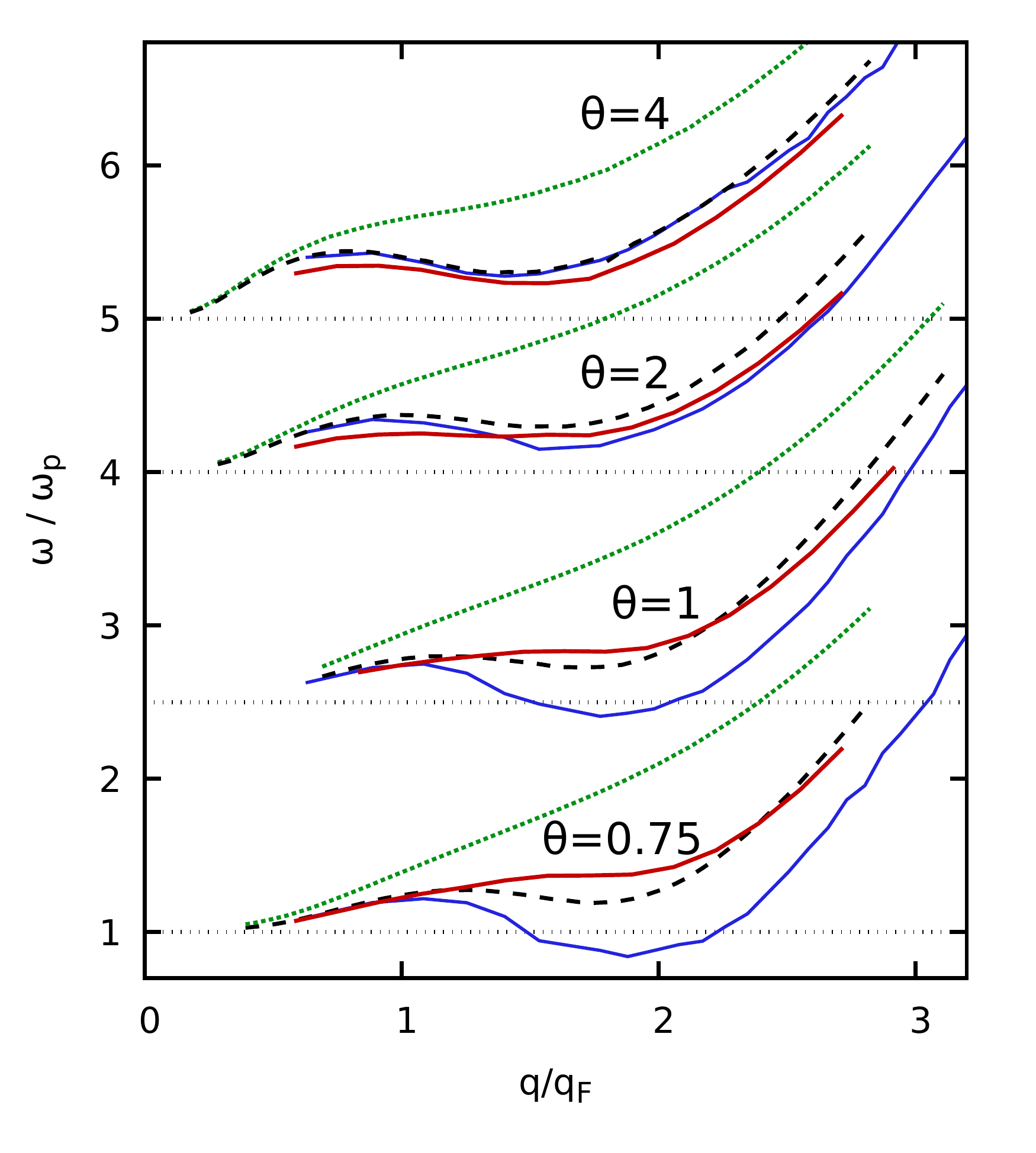}
\caption{\label{fig:overview}
Wavenumber dependence of the spectrum of density fluctuations $\omega(q)$ at different conditions. Dotted green: RPA; dashed black: \emph{static approximation}; solid blue: exact PIMC results~\cite{dornheim_dynamic}; solid red: new \emph{electronic pair alignment} model [Eq.~(\ref{eq:dispersion})] introduced in this work. Left: results for different value of the density parameter $r_s$ at $\theta=1$; right: results for different values of the temperature parameter $\theta=k_\textnormal{B}T/E_\textnormal{F}$ for $r_s=10$.
}
\end{figure*}

\subsection*{Spectral representation of the DSF}

An additional motivation for the new \emph{electronic pair alignment} model is given by the exact spectral representation of the DSF~\cite{quantum_theory},
\begin{eqnarray}\label{eq:spectral}
S(\mathbf{q},\omega) = \sum_{m,l} P_m \left\|{n}_{ml}(\mathbf{q}) \right\|^2 \delta(\omega - \omega_{lm})\ .
\end{eqnarray}
Here $l$ and $m$ denote the eigenstates of the full $N$-body Hamiltonian, $\omega_{lm}=(E_l-E_m)/\hbar$ is the energy difference, and $n_{ml}$ is the usual transition element from state $m$ to $l$ induced by the density operator $\hat{n}(\mathbf{q})$. Evidently, Eq.~(\ref{eq:spectral}) implies that $S(\mathbf{q},\omega)$ is fully defined by the possible transitions between the (time-independent) eigenstates; the full frequency dependence comes from the corresponding energy differences. In other words, no time propagation is needed.
The translation of our \emph{electronic pair alignment} model and the corresponding impact of $\Delta W(q)$ on $\omega(q)$ into the language of Eq.~(\ref{eq:spectral}) is then straightforward. Firstly, we assume a continuous distribution of eigenstates, which we examine in coordinate space. In the regime of $\lambda\sim d$, the excitations primarily involve only two particles, as the effective potential $\phi(r)$ decays rapidly with $r$. The probability $P(r)=n g(r)$ thus plays the role of $P_m$ in Eq.~(\ref{eq:spectral}), and the energy shift can be expressed as $\omega_{lm} = \Delta W_{lm} + \Delta K_{lm}$.
The kinetic contribution is accurately captured by the RPA as we demonstrate in the next section.
Finally, we note that Eq.~(\ref{eq:spectral}) even gives us some insight into the nontrivial shape of the exact PIMC results for $S(\mathbf{q},\omega)$ shown in Fig.~\ref{fig:pair} in the main text. In particular, the roton peak around $\omega_\textnormal{p}$ must be due to transitions where the down-shift $\Delta W$ is substantial. This is only the case for electron pairs that have been separated by $r<r_s$ in the initial state. The substantial reduction in the interaction energy of such a pair due to an excitation with $\lambda\sim d$ is thus the microscopic explanation for the observed \emph{roton feature}.

\subsection*{Momentum distribution of the correlated electron gas}

In Fig.~\ref{fig:single-particle}, we show the momentum distribution function $n(k)$ of the UEG at the electronic Fermi temperature $\theta=1$. Specifically, the symbols show exact PIMC results~\cite{Dornheim_PRB_nk_2021} for different values of $r_s$, and the dashed black line the Fermi distribution describing a noninteracting Fermi gas at the same conditions. Evidently, $n(k)$ is hardly influenced by the Coulomb interaction for both $r_s=4$ (blue diamonds) and $r_s=10$ (green crosses); correlation effects only manifest for much larger $r_s$, cf.~the yellow triangles that have been obtained for a strongly coupled electron liquid ($r_s=50$). This is a strong indication that the main error in $\omega_\textnormal{RPA}(q)$ is due to $\Delta W$ and not the kinetic part.

\subsection*{Results for other temperatures and densities}\label{sec:supplement3}

In the main text, we have restricted ourselves to the representative cases of $r_s=4$ (metallic density) and $r_s=10$ (boundary to the electron liquid regime~\cite{review}) at the electronic Fermi temperature, $\theta=k_\textnormal{B}T/E_\textnormal{F}$. The validity of our \emph{electronic pair alignment} model is demonstrated for a vast range of densities (left) and temperatures (right) in Fig.~\ref{fig:overview}.

\subsection*{Connection of the electronic pair alignment model to Green functions theory} 
In the following, we connect the shift of the plasmon dispersion to the energy change of a test particle, $\Delta W_{xc}$. In  Ref.~\cite{kwong_prl-00}, a direct relation between the DSF and the single-particle nonequilibrium Green function (NEGF) $\delta G^<$ was derived that is produced by a short monochromatic field pulse, $U(t,q)=U_0(t)\cos{q x}$, and is valid in case of a weak excitation (linear response). Here we rewrite this in terms of the spectral function of the occupied states, $\delta A$,
\begin{align}
    S(\omega,q) &= \frac{1}{\pi n_0 \hbar U_0(\omega)}\sum_p  \,\delta A(p,\omega,U)\,,
\label{eq:dsf-gless}\\
\delta A(p,\omega,U) &=  A(p,\omega,U) -  A(p,\omega,0)\,,
\end{align}
\label{eq:delta-a-def}
where the argument $U$ comprises the dependencies on $U_0$ and $q$. Note that $\delta A$ is proportional to $U_0$, cancelling its appearance in the denominator. Thus, in linear response there is a direct linear relation between the frequency dependencies of the DSF and the field-induced correction to the single-particle spectral function. Now the question is how the peak position of the DSF, $\omega(q)$, that is discussed in the main part of the paper, is related to the peak position $\delta E(p,U)$ of the spectral function $\delta A$. 

To answer this question we follow the approach of Ref.~\cite{kwong_prl-00}  and outline the main steps. First the Keldysh-Kadanoff-Baym equations (KBE) are solved for the NEGF, $G(t_1,t_2)$, in the two-time plane, in the presence of the field $U$. The spectral information is then contained in the dependence of $A(p,\tau,U)$ on the difference time, $\tau=t_1-t_2$, and the numerical result can be expressed as a Fourier series
\begin{align}
    A(p,\tau,U) & = \sum_a C_a\, e^{i\frac{E_a(p)}{\hbar} \tau}e^{-\Gamma_a(p) \tau} \,.
    \label{eq:a-ansatz}
\end{align}
The exponential damping ansatz is known to be a poor approximation for small $\tau$, and can be straightforwadly improved; 
however, for the present discussion ansatz (\ref{eq:a-ansatz}) is sufficient.

In the field-free case, $U_0 \to 0$, and a given exchange-correlation selfenergy of the uniform electron gas, $\Sigma_{xc}(p,\tau)$ [the Hartree term vanishes for the UEG], the sum (\ref{eq:a-ansatz}) contains only a small number $s$ of terms. For example, in the quasiparticle approximation, there is only one term, $s=1$, with $E_1(p)=\frac{p^2}{2m} + \rm Re\, \Sigma_{xc}(p,\tau)$ and $\hbar\Gamma_1(p)= \rm Im \, \Sigma_{xc}(p,\tau)$. Since the system is stationary, there is no dependence on the center of mass time, $T=(t_1+t_2)/2$. Now, when the field $U$ is turned on, it excites plasma oscillations with wave number $q$ which give rise to one additional contribution to the sum with (within linear response) 
\begin{align}
    E_2(p,q) &= \delta \Sigma_H(p,q) + \rm Re\, \delta\Sigma_{xc}(p,q)\,,\\
    \hbar\Gamma_2(p,q) &= \rm Im\, \delta\Sigma_{xc}(p,q)\,,
    \label{eq:gamma2}
\end{align}
where $\delta \Sigma_H$ and $\delta\Sigma_{xc}$ are the linear perturbations of the selfenergies due to the external field, where we suppress the time dependencies, see Ref.~\cite{bo1999} for more details.
The specific contribution to the single-particle spectrum that is caused by plasmons can be isolated by considering the difference $\delta A$, where $s$ field-free terms cancel. Now, Fourier transforming with respect to $\tau$ yields a Lorentzian in frequency space with the peak position of $\delta A(p,\omega)$ given by $E_2(p,q)=\delta E(p,q)$,
\begin{align}
   \delta E(p,q) & \approx \delta \Sigma_H(p,q) +  {\rm Re}\,\delta \Sigma_{xc}(p,q)\,,
   \label{eq:delta-e}
\end{align}
with the life time $\Gamma_2$.
If the KBE are solved in the presence of the field on the mean field level ($\Sigma_{xc}=0$), the second term in Eq.~(\ref{eq:delta-e}) vanishes which is known to yield the plasmon spectrum (peak of the DSF) on RPA-level, $\omega_{\rm RPA}(q)$ \cite{kwong_prl-00}. If exchange-correlation effects are restored, the plasmon spectrum and energy spectrum $\delta E$
undergo synchronous changes,
\begin{align}
\omega_{\rm RPA}(q) &\to \omega(q) \nonumber\\
\delta \Sigma_H(p,q) &\to \delta \Sigma_H(p,q) +  {\rm Re}\,\delta \Sigma_{xc}(p,q)\,.
\nonumber
\end{align}
Equivalently, we may subtract the terms on the left-hand side. This yields, on one hand, the frequency change, $\Delta \omega_{\rm xc}= \omega(q)-\omega_{\rm RPA}(q)$, that is caused by exchange-correlation effects. On the other hand, this leads to the  xc-induced difference of energy dispersions
\begin{align}
    \Delta \delta E_{\rm xc}(p,q) 
    & \approx  {\rm Re}\,\delta \Sigma_{\rm xc}(p,q)\,.
\label{eq:deltadelta_exc}
\end{align}
Thus, we have established a direct link between the two exchange correlation energy effects, $\Delta \omega_{\rm xc}$ and ${\rm Re}\,\delta \Sigma_{xc}(p,q)$. Taking into account the linear relation (\ref{eq:dsf-gless}) and subtracting the mean field (RPA) expressions, we expect a proportionality also for the peak positions, 
\begin{align}
\Delta \omega_{\rm xc}(q) \sim \sum_p {\rm Re}\,\delta \Sigma_{\rm xc}(p,q)
\label{eq:dom-dsig-xc}
\end{align}

While, the KBE procedure has been successfully demonstrated for the computation of the plasmon spectrum in Ref.~\cite{kwong_prl-00}, the change of the single-particle energy, Eq.~(\ref{eq:deltadelta_exc}), is presently not available.
Physically, $\delta \Sigma_{xc}(p,q)$ has the meaning of the field-induced change of the energy of a test particle that is related to its interaction with the medium \cite{kremp_book}.
 Since $\Delta W_{xc}$ has exactly this meaning (an approximation to it) we  conclude that 
\begin{align}
\sum_p\Delta \delta E_{\rm xc}(p,q) \sim \Delta W_{\rm xc}(q)\,.
\end{align}
%
Together with the proportionality (\ref{eq:dom-dsig-xc}) this gives a connection to Eq.~(\ref{eq:dispersion}) of the main text.

\section*{Acknowledgments}

This work was partly funded by the Center for Advanced Systems Understanding (CASUS) which is financed by Germany's Federal Ministry of Education and Research (BMBF) and by the Saxon Ministry for Science, Culture and Tourism (SMWK) with tax funds on the basis of the budget approved by the Saxon State Parliament. MB acknowledges support by the DFG via project BO1366/15.
The PIMC calculations were carried out at the Norddeutscher Verbund f\"ur Hoch- und H\"ochstleistungsrechnen (HLRN) under grant shp00026 and on a Bull Cluster at the Center for Information Services and High Performance Computing (ZIH) at Technische Universit\"at Dresden.

\bibliography{bibliography}
\end{document}